\documentclass[]{spie}  

\usepackage{amsmath,amsfonts,amssymb}
\usepackage{graphicx}
\usepackage{float}
\usepackage{caption}
\usepackage{subcaption}
\usepackage{cite}
\graphicspath{{images/}}
\usepackage{xcolor}
\usepackage[colorlinks=true, allcolors=blue]{hyperref}

\title{Design of Novel Loss Functions for Deep Learning in X-ray CT}

\author[a]{Obaidullah Rahman}
\author[a]{Ken D. Sauer}
\author[b]{Madhuri Nagare}
\author[b]{Charles A. Bouman}
\author[c]{Roman Melnyk}
\author[c]{Jie Tang}
\author[c]{Brian Nett}

\affil[a]{Department of Electrical Engineering, University of Notre Dame, U.S.A.}
\affil[b]{School of Electrical and Computer Engineering, Purdue University, U.S.A.}
\affil[c]{General Electric Healthcare, Waukesha, U.S.A.}

\authorinfo{Further author information: (Send correspondence to O.R.)\\O.R.: E-mail: orahman@nd.edu}

\begin{document} 
\maketitle

\begin{abstract}
Deep learning (DL) shows promise of advantages over conventional signal processing techniques in a variety of imaging applications.  The networks’ being trained from examples of data rather than explicitly designed allows them to learn signal and noise characteristics to most effectively construct a mapping from corrupted data to higher quality representations.  In inverse problems, one has options of applying DL in the domain of the originally captured data, in the transformed domain of the desired final representation, or both.

X-ray computed tomography (CT), one of the most valuable tools in medical diagnostics, is already being improved by DL methods.  Whether for removal of common quantum noise resulting from the Poisson-distributed photon counts, or for reduction of the ill effects of metal implants on image quality, researchers have begun employing DL widely in CT. The selection of training data is driven quite directly by the corruption on which the focus lies.  However, the way in which differences between the target signal and measured data is penalized in training generally follows conventional, pointwise loss functions.

This work introduces a creative technique for favoring reconstruction characteristics that are not well described by norms such as mean-squared or mean-absolute error.  Particularly in a field such as X-ray CT, where radiologists’ subjective preferences in image characteristics are key to acceptance, it may be desirable to penalize differences in DL more creatively.  This penalty may be applied in the data domain, here the CT sinogram, or in the reconstructed image.  We design loss functions for both shaping and selectively preserving frequency content of the signal.

\end{abstract}

\keywords{Deep learning, neural network, X-ray CT, novel loss functions, spectral shaping.}

\section{INTRODUCTION}
\label{sec:intro}  

\begin{figure}[h]
\centering
\includegraphics[width = 3.0in]{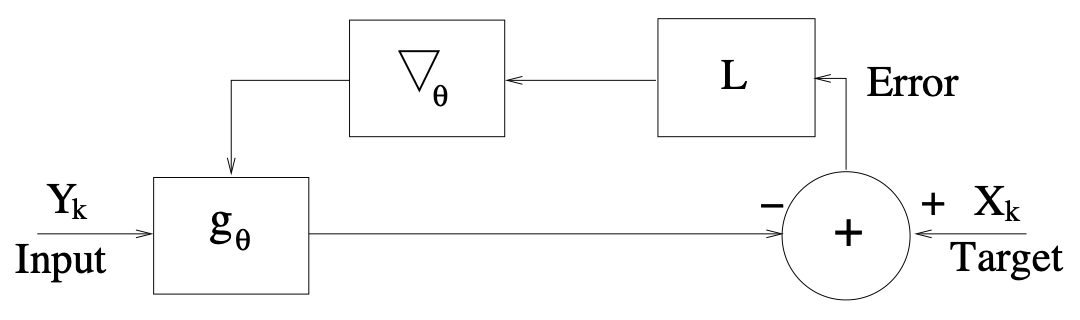}
\caption{\label{fig:DL_Arch} Training of neural network. Parameters governing system behavior are denoted by $\theta$. The gradient of the loss function’s penalization of error ($L$), as a function of $\theta$, is used to improve the averaged match between target and output of network during training.}
\end{figure}

Artificial neural networks (ANN) have been increasingly finding success in X-ray computed tomography (CT) \cite{ananthabhotla2019towards, yang2018low, suzuki2018transforming, yang2017ct, ghani2018cnn, lee2019deep, yuan2018sipid, wang2021tomographic, thibault2021image, rahman2021mbir}. 
ANN in imaging are designed by adjusting strengths of interconnections among artificial neurons with the goal of making the network’s output, on the average, as close as possible to the ideal form of the image. 
This ideal form may be well known in training phase of the ANN, in which one may start with a perfect signal as the “target” and then corrupt it according to the character of noises and artifacts typically encountered in application. 
Alternatively, the target image may be imperfect, but far less afflicted with error than those encountered as measurements. 
In training, simple multiplicative coefficients or other representations of neural interconnections are iteratively adjusted to minimize some average measured error, or loss, between an ensemble of network-processed input data and their respective target images, as represented in Figure \ref{fig:DL_Arch}.
The measured loss is backpropagated through the ANN to provide gradients to correct the connections and reduce loss, thus “learning” the inverse operator. Following training, the network may be applied to new data sets in order to reduce their content of error as described by the system’s loss function. The process is, with increasing frequency, titled “deep learning” because more powerful computational resources have allowed more layers in the ANN, hence a “deeper” network.

Probably the most common loss function applied has been mean-squared error. 
Let us define $Y$ as the input data, which we model as a function of some ideal, target image $X$, or $Y =h(X)$. 
The task of the ANN is to extract from $Y$ a rendering close to the unknown, ideal image.
 If we define $g = h^{-1}$, our training would seek to learn $g$ to produce $X = g(Y)$. 
 Equality is seldom achievable due to noise or other corruption, and we optimize in the sense of average, possibly weighted, error. 
 If we use the variable $k$ to index among training pairs, $n$ to index entries in vectors $X_k$ and $Y_k$, and $\theta$ to represent the variable parameters of the ANN, our DL-trained mapping $g_\theta$ for the mean-squared error case may be expressed in terms of
\begin{eqnarray}
\theta = \underset{\theta}{\operatorname{\textit{argmin}}} \ \sum_{k,n} w_{k,n}[X_{k,n} - (g_\theta (Y_k))_n]^2
\end{eqnarray}
in which the weightings $w_{k,n}$ may be fixed in either or both variables, or may be adapted according to relative local characteristics of data. 
This weighted, mean-squared penalty on the standard error, $S_k \triangleq X_k-g_\theta(Y_k)$, has a number of potential advantages, including being statistically well-matched to Gaussian noise. 
In cases where less severe penalization of large errors is desired, squared error may be replaced by absolute error, similarly to penalty adjustment in edge-preserving regularization.

While simple norms such as expressed above provide highly useful loss metrics, it has long been recognized in the image processing community that they may be less than ideal for applications in which the final receiver for the system’s output is a human observer. 
Various metrics for perceptual loss have been designed in hopes of optimizing the elusive human-interpreted quality of audio \cite{ananthabhotla2019towards} and visual data \cite{yang2018low}. 
For diagnostic CT imaging, in which much analysis is performed by radiologists, more subjective quality metrics are applied by the end users of the technology, and spectral content of residual noise, plateauing of image levels in low-contrast areas and other context-dependent evaluations must be addressed.

This work consists of a novel class of loss metrics which may expand the usefulness of DL in X-ray CT. 
We generalize the sense of optimality to
\begin{eqnarray}
\hat{\theta} = \underset{\theta}{\operatorname{\textit{argmin}}} \ \sum_k L [X_k, Y_k, g_\theta(Y_k)],
\end{eqnarray}
where $L$ is now a function that may capture any number of spectral and spatial characteristics in the error. 
In the X-ray CT arena, we may choose to improve the signal in either the sinogram domain, where measurements are made directly, or in the image domain after reconstruction by any existing algorithm. 
The signal and error statistics in these two differ, leading to designs tailored for each case. 
In the following, we describe one embodiment of the design.

\section{METHOD}

Conventional, point-wise mean-squared error as loss
may be thought of as a flat spectral penalty. 
However, in cases where we wish to focus on removing artifacts with low or medium spatial frequency content, penalizing all frequencies equally may be counter-productive.
Given that many well-developed, edge-preserving techniques are available for removing high-frequency noise, particularly in the image domain, low-signal correction in CT may in some cases be better served by training the network to remove errors only in lower frequencies. 
In this case, we propose a loss function $L$ in Eq.\ (\ref{eq:Spectral_shaping_equation}) that may take the form
\begin{eqnarray}
L[S_k ] \triangleq \phi[f_1(S_k )],\label{eq:Spectral_shaping_equation}
\end{eqnarray}
where $\phi$ is a suitable error metric applied only within the passband of the lowpass filter $f_1$. 
The higher frequency error becomes a ``don’t care” element for the network. 
Alternatively, band-pass or high-pass filtering may focus loss on those portions of the error spectrum. 
Particularly in three-dimensional image vectors, frequencies may be treated differently along the three axes. 
This forms the first part of our novel loss function. 

The discussion above is most commonly addressed to conventional CT imagery in two or three dimensions, in which spatial frequency has roughly equivalent meaning in all dimensions.
However, the present methods are intended at least as importantly for use in the native domain of the data, the sinogram. 
Application of the type of loss function in Eq.\ (\ref{eq:Spectral_shaping_equation}) in the sinogram requires modeling behavior in such coordinates as row, channel and view, where the first two index in the detector panel of the CT gantry, and the last indexes the distinct rotating, two-dimensional views of patient or object. 
In this case, the error filtering operation will need to be spatially adapted, as statistics of both the underlying signal and the corrupting noise vary spatially in the sinogram domain.

It has been widely observed in the DL community that networks appear to have a strong tendency toward elimination of high frequencies in the output and this may occur even when the penalized loss is restricted to low frequency error
as in Eq.\ (\ref{eq:Spectral_shaping_equation}).
An example application is using DL for low signal correction, where some of the most problematic artifacts are of low to medium spatial frequency. 
Here, it may be advantageous to retain parts of the error spectrum in the output when the correction network is applied in the sinogram domain. 
Powerful, adaptive denoisers in the image domain can capitalize on the relatively stationary underlying image statistics to remove higher frequency noise with little damage to edge resolution. 
Thus, we may wish to actively discourage suppression of this part of the error signal in the first stage of processing
in order to preserve both resolution and desirable texture.
We propose a second part of the loss function that will penalize removal of components of the signal $Y_k$ according to their spectral content.
This component of the loss may be expressed similarly to Eq.\ (\ref{eq:Spectral_shaping_equation}), but with the argument redefined as
\begin{eqnarray}
T_k \triangleq Y_k-g_\theta(Y_k) \label{eq:Input_output_error} \\
L[S_k,\ T_k ] \triangleq \phi[f_1(S_k ) +\alpha f_2(T_k)],\label{eq:Preservation_error_equation}
\end{eqnarray}
An realization of the system is shown below in Figure \ref{fig:DL_Arch_Inp_Pres}. 
It includes the two loss functions discussed previously. 
The first loss, realized by the right branch, penalizes the error from Eq.\ (\ref{eq:Spectral_shaping_equation}) filtered by $f_1$. 
The left branch features the error from Eq.\ (\ref{eq:Input_output_error}), where a different portion of the spatial frequency spectrum of error within the passband of $f_2$ is penalized. 
The two types of error signals are combined before the application of the norm $\phi$ and the gradient for backpropagation. 
The weighting factor $\alpha$ could be any positive value, with increase resulting in more of the desired frequency components preserved in the output.
\begin{figure}[h]
\centering
\includegraphics[width = 2.5in]{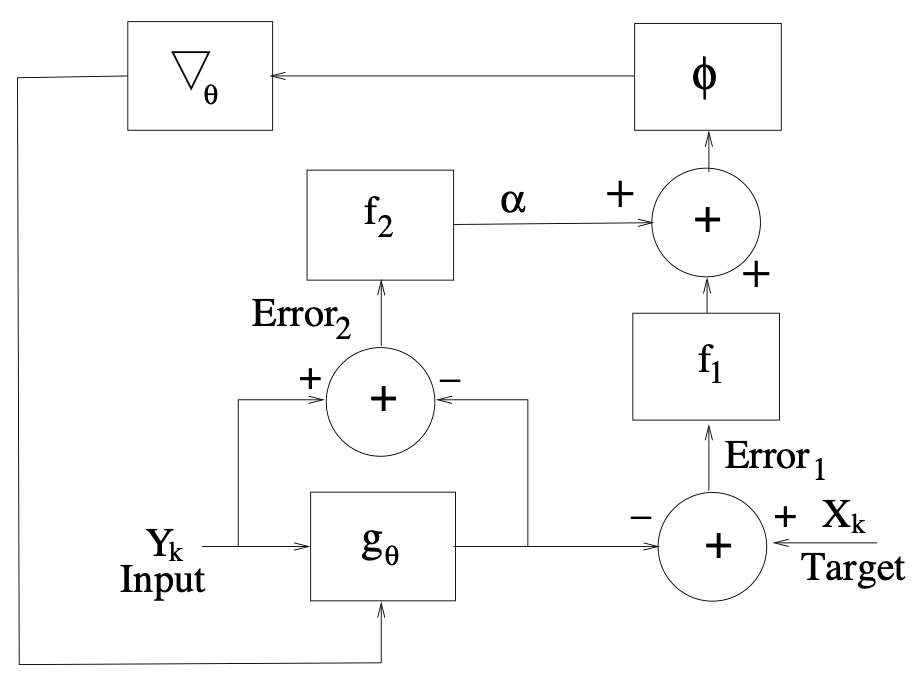}
\caption{\label{fig:DL_Arch_Inp_Pres}Training of system to encourage the output to mimic the target content as selected by filter $f_1$, but refrain from removal of input signal content as selected by filter $f_2$}.
\end{figure}
The responses of filters $f_1$ and $f_2$ plus the parameter $\alpha$ appear to provide a great deal of control
over the inference behavior of the network.
In an extremely conservative case, with $f_1 = f_2 = 1.0\ \forall \ \omega$ and $\alpha = 1$, 
the composite error becomes 
\begin{eqnarray}
X_k + Y_k - 2g_\theta (Y_k),
\end{eqnarray}
which will simply place the optimum output midway between the target and the input.

\section{RESULTS}
\label{sec:result}

\begin{figure}[h]
\centering
\includegraphics[width = 2.5in]{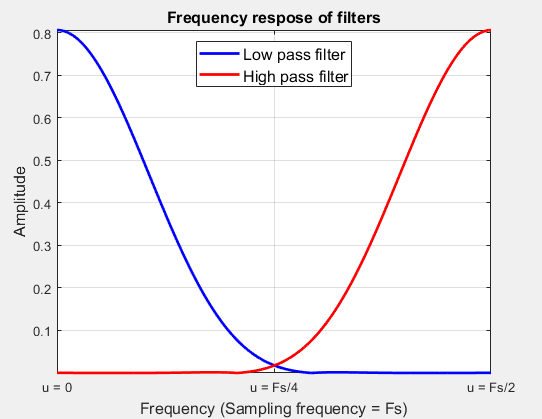}
\caption{\label{fig:filters}Filters used. The LP filter is $f_1$ and the HP filter is $f_2$}
\end{figure}

\begin{figure}[ht]
\centering
\subfloat[][]{\includegraphics[width=\columnwidth]{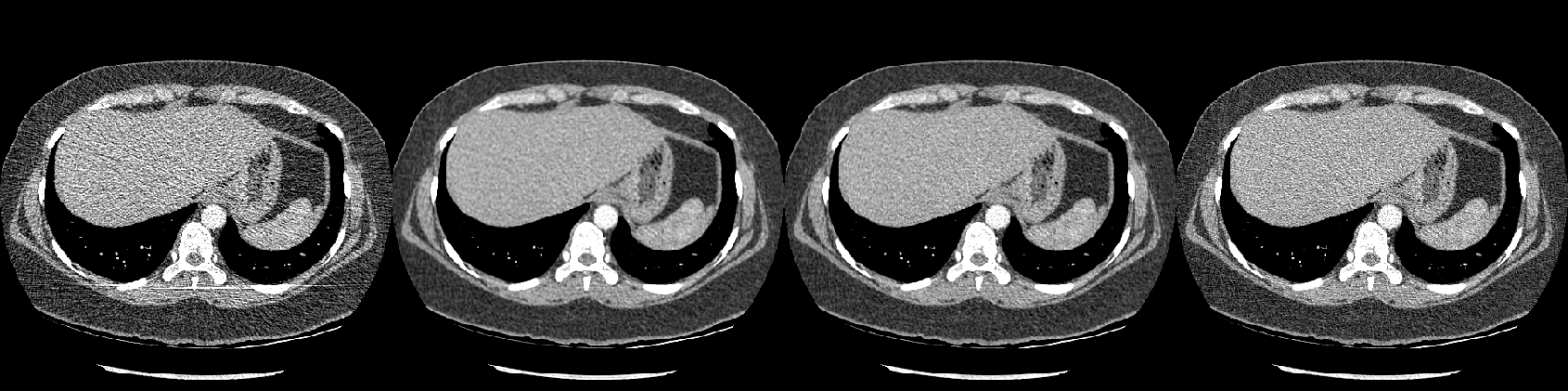}}
\hspace{0.1 cm}
\subfloat[][]{\includegraphics[width=\columnwidth]{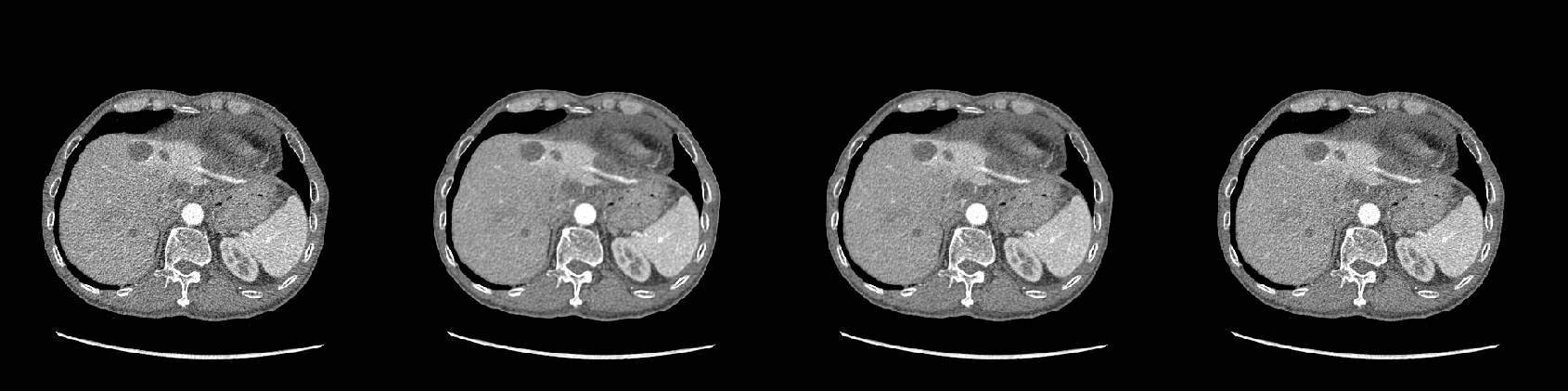}}
\caption{\label{fig:Input_Preserve}{Reconstructed image (Left to right) Uncorrected;  corrected with low pass filter loss ($\alpha=0$); $\alpha = 0.6$; $\alpha = 0.8$.}}
\end{figure}

Parts of this method have been preliminarily tested with phantom and clinical data. Below are a few results with the latter.
In this configuration, the training loss was the weighted sum of low pass (LP) filtered error between output and target, and high pass (HP) filtered error between input and output. 
The filters are shown in the Figure \ref{fig:filters}. 
The DL network was trained to operate in the original domain i.e. counts domains. 
Training data consisted of high-dosage Kyoto phantom scans as targets,
with synthetic photon counting and electronic noise added to form input sinograms.
We can see in Figure \ref{fig:Input_Preserve} the increase in fine-grain texture i.e. high frequency components in the output with increase in $\alpha$.

\begin{figure}[h]
\centering
\includegraphics[width = 3.0in]{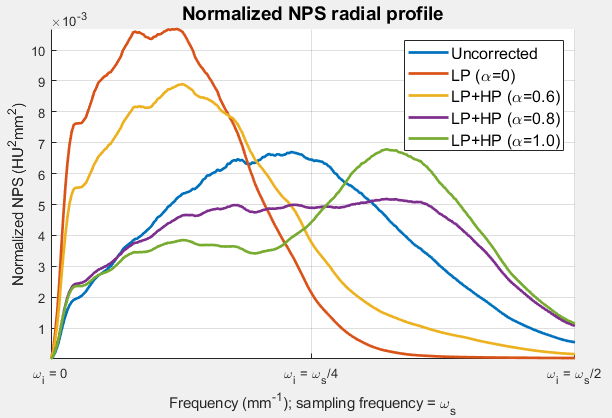}
\caption{\label{fig:NPS_Curve}Normalized NPS curves}
\end{figure}

The noise power spectra (NPS), shown in Figure \ref{fig:NPS_Curve}, were measured in the liver region of reconstructed clinical images. 
The NPS resulting from the use of only low pass in the loss function ($\alpha = 0$) can be seen to lack much
high frequency content. 
Use of the high pass filter on the error between the input and the output preserves some of the high frequency components,
retaining resolution along with high-frequency noise.
The value of $\alpha$ can be adjusted based on the balance between NPS qualities and noise tolerance in the image.
To assess the flatness of the NPS curve, entropy measurement was performed as
\begin{eqnarray}
Entropy =  \sum_{\omega_i=0}^{\omega_s/2} NPS(\omega_i) log_2 \frac{1}{NPS(\omega_i)},
\end{eqnarray}
where $\omega_i$ is the discrete spatial frequency and $\omega_s$ is the spatial sampling frequency. 
It can be seen in Table \ref{tab:NPS_Metric} that the flatness of the NPS increases with $\alpha$ as far as $0.8$, but it suffers from excessive high frequency emphasis for $\alpha$ of $1.0$. This case exhibits undesirable streaks in the image as well.
\begin{table}[h]
\begin{center}
\begin{tabular}{|c|c|c|c|c|c|}
\hline
\multicolumn{6}{|c|}{Flatness metric  (entropy in bits of information)}\\
\hline
Ideal case & Uncorrected & $\alpha=0$ & $\alpha=0.6$ & $\alpha=0.8$ & $\alpha=1.0$\\
\hline
8.00 & 7.76 & 7.07 & 7.46 & 7.90 & 7.87\\
\hline
\end{tabular}
\caption{\label{tab:NPS_Metric} Entropy as a measurement of flatness of NPS curves. Higher value indicates flatter, more desirable NPS}
\end{center}
\end{table}
\section{CONCLUSION}

This paper presents a combination of two frequency-weighted loss function components for a deep network,
furnishing potentially better control of the behavior of the network in removing signal corruption.
The first part of the DL loss function employed here restricts training loss to lower frequency error between a target data set and
the input set processed by the network.
The second component of the loss ensures the preservation of select error content from the uncorrected data, with the intent
of delegating any removal of that error to a later stage of processing.
This results in network's ability  to retain desired traits in the data according to chosen models for training loss. 
In our example application, improvement in the texture of the reconstructed image was observed and confirmed with the NPS metric. 
Further work will test the value of this design in improving the noise/resolution trade-off
in the presence of image-domain postprocessing.
We have developed this novel DL loss function design for X-ray CT imaging, but it can easily find application in other areas.

\bibliography{report} 
\bibliographystyle{spiebib} 

\end{document}